\documentclass[ALICE,manyauthors]{cernphprep}

\usepackage{amsmath}
\usepackage{latexsym}
\usepackage{color}
\usepackage{ulem}
\usepackage{graphicx}
\usepackage{graphics}
\usepackage{epsfig}
\usepackage{amssymb}
\usepackage{units}
\usepackage{cite}
\usepackage[switch]{lineno}
\usepackage{calc}
\usepackage{endnotes}
\usepackage[colorlinks,citecolor=blue,urlcolor=blue,linkcolor=blue]{hyperref}

\begin{document}%
%
%
\begin{titlepage}
\PHnumber{2012-136}                 
\PHdate{18 May 2012}              
%
%
\title{Transverse sphericity of primary charged particles in minimum bias proton-proton 
collisions at $\mathbf{\sqrt{s}=0.9}$, $\mathbf{2.76}$ and $\mathbf{7}$ TeV}
\ShortTitle{Transverse sphericity of primary charged particles in MB proton-proton collisions}   
%
\Collaboration{ALICE Collaboration%
         \thanks{See Appendix~\ref{app:collab} for the list of collaboration
                      members}}
\ShortAuthor{ALICE Collaboration}      

\begin{abstract}
 Measurements of the sphericity of primary charged particles 
in minimum bias proton--proton collisions at $\sqrt{s}=0.9$, $2.76$ and $7$ TeV with the ALICE 
detector at the LHC are presented. The observable is linearized to be collinear safe and is measured 
in the plane perpendicular to the beam direction using primary charged tracks with $p_{\rm T}\geq0.5$ 
GeV/$c$ in $|\eta|\leq0.8$. 
The mean sphericity as a function of the charged particle multiplicity at mid-rapidity ($N_{\rm ch}$) 
is reported for events with different $p_{\rm T}$ scales (``soft'' and ``hard'') defined by the transverse momentum 
of the leading particle. 
In addition, the mean charged particle transverse momentum versus multiplicity is presented for the 
different event classes, and the sphericity distributions in bins of multiplicity are presented.  
The data are compared with calculations of standard Monte Carlo event generators. 
The transverse sphericity is found to grow with multiplicity at all collision energies, 
with a steeper rise at low $N_{\rm ch}$, whereas the event generators show the opposite tendency.
The combined study of the sphericity and the mean $p_{\rm T}$ with multiplicity indicates 
that most of the tested event generators produce events with higher multiplicity by generating more back-to-back jets resulting 
in decreased sphericity (and isotropy). The PYTHIA6 generator with tune PERUGIA-2011 exhibits a noticeable improvement in describing the data, compared to the other tested generators.
\end{abstract}


\end{titlepage}
\setcounter{page}{2}

\section{Introduction}
Minimum bias proton--proton collisions present an interesting, and theoretically, challenging subject 
for detailed studies. Their understanding is important for the interpretation of measurements of heavy-ion collisions, and in the search for signatures of new physics at the Large Hadron Collider~(LHC) and Fermilab. 
However, the wealth of experimental information is currently poorly understood by theoretical models or 
Monte Carlo (MC) event generators, which are unable to explain with one set of parameters all the 
measured observables. Examples of measured observables which are not presently well described theoretically
include the reported multiplicity distribution~\cite{aliceNch1,aliceNch2,atlas:nch}, the transverse momentum distribution~\cite{cms:pt} and the variation of the transverse momentum with multiplicity~\cite{cms:ptnch,atlas:ptnch,alice30}.

In this paper, we present measurements of the transverse sphericity for pp minimum bias events over a wide multiplicity range  at several energies using the ALICE detectors. Transverse sphericity is a momentum space variable, commonly classified as an event shape observable~\cite{sphericity:1}. 
Event shape analyses, well known from lepton collisions~\cite{sphericity:2,sphericity:3,sphericity:4}, also offer interesting possibilities in hadronic collisions, 
such as the study of hadronization effects, underlying event characterization and comparison of pQCD computations 
with measurements in high $E_{\rm T}$ jet events~\cite{banfi2004,cmsES,cdfESII}.

The goal of this analysis is to understand the interplay between the event shape, the charged particles multiplicity, and their transverse momentum distribution; hence, the present paper is focused on the following aspects:
\begin{itemize}
\item{} The evolution of the mean transverse sphericity with multiplicity for different subsets of events 
defined by the transverse momentum of the leading particle;
\item{} the behavior of the mean transverse momentum as a function of multiplicity;
\item{} the normalized transverse sphericity distributions for various multiplicity ranges.
\end{itemize}

The results of these analyses are compared with event generators and will serve for a better understanding of the underlying processes in proton-proton interactions at the LHC energies.

\section{Event shape analysis}
At hadron colliders, event shape analyses are restricted to the transverse plane in order to avoid the bias 
from the boost along the beam axis~\cite{banfi2004}. 
The transverse sphericity is defined in terms of the
eigenvalues: $\lambda_{1}>\lambda_{2}$ of the transverse momentum matrix: 

\begin{displaymath}
\mathbf{S_{xy}^{Q}} = \frac{1}{\sum_{i} {p_{\rm T}}_{i}}\sum_{i}
	\left(\begin{array}{cc}
{p_{\rm x}}_{i}^{2}      &  {p_{\rm x}}_{i} \, {p_{\rm y}}_{i} \\
{p_{\rm y}}_{i} \, {p_{\rm x}}_{i} &  {p_{\rm y}}_{i}^{2} 
	\end{array} \right)
\end{displaymath}

where $({p_{\rm x}}_{i},{p_{\rm y}}_{i})$ are the projections of the transverse momentum of 
the particle $i$.

Since $\mathbf{S_{xy}^{Q}}$ is quadratic in particle momenta, this sphericity is a non-collinear safe quantity in pQCD. For instance, if a hard momentum along the $x$ direction splits into two equal collinear momenta, then the sum $\sum_{i} {p_{\rm x}}_{i}^2$ will be half that of the original momentum. To avoid this dependence on possible collinear splittings, the transverse momentum matrix is linearized as follows:

\begin{displaymath}
\mathbf{S_{xy}^{L}} = \frac{1}{\sum_{i} {p_{\rm T}}_{i}}\sum_{i}
	\frac{1}{ {p_{\rm T}}_{i} }\left(\begin{array}{cc}
{p_{\rm x}}_{i}^{2}      &  {p_{\rm x}}_{i} \, {p_{\rm y}}_{i} \\
{p_{\rm y}}_{i} \, {p_{\rm x}}_{i} &  {p_{\rm y}}_{i}^{2} 
	\end{array} \right)
\end{displaymath}
The transverse sphericity is defined as
\begin{equation}
S_{\rm T} \equiv \frac{2\lambda_{2}}{\lambda_{2}+\lambda_{1}}\,.
\end{equation}
By construction, the limits of the variable are related to specific configurations in the transverse plane
\begin{displaymath}
S_{\rm T}=\left(\lbrace \begin{array}{ll}
=0 & \textrm{``pencil-like'' limit} \\
=1 & \textrm{``isotropic'' limit}
\end{array} \right. \,.
\end{displaymath}
This definition is inherently multiplicity dependent, for instance, $S_{\rm T}\rightarrow 0$ 
for very low multiplicity events.

\section{Experimental conditions}
The relevant detectors used in the present analysis are the Time Projection Chamber (TPC) and the Inner Tracking 
System (ITS), which are located in the central barrel of ALICE inside a large solenoidal magnet 
providing a uniform $0.5$ T field~\cite{alice-ppr1}. 

The ALICE TPC is a large cylindrical drift detector with a central membrane maintained at -100 kV 
and two readout planes at the end-caps composed of 72 multi-wire proportional chambers~\cite{tpc-ref}. The active volume is limited to $85 < r < 247$ cm and 
$-250 < z < 250$ cm in the radial and longitudinal directions, respectively. The material budget between the interaction point and the active volume of the TPC corresponds to $11\%$ of a radiation length, averaged in $|\eta|\leq0.8$. The central membrane divides the nearly 90 $m^{3}$ active volume into two halves. The homogeneous drift field of 400 V/cm in the Ne-CO$^{2}$-N$^{2}$ (85.7$\%$-9.5$\%$-4.8$\%$) gas
mixture leads to a maximum drift time of 94 $\mu$s. The typical gas gain is $10^{4}$~\cite{alice30}. 

The ITS is composed of high resolution silicon tracking detectors, arranged in six cylindrical layers at radial 
distances to the beam line from 3.9 to 43 cm. The two innermost layers are Silicon Pixel Detectors (SPD), 
covering the pseudorapidity ranges $|\eta|<$2 and $|\eta|<$1.4, respectively.
A total of 9.8 millions $50\times425$ $\mu$m$^{2}$ pixels enable the reconstruction of the primary event vertex 
and the track impact parameters with high precision. The SPD was also included in the trigger scheme for data 
collection. The outer third and fourth layers are formed by Silicon Drift Detectors (SDD) with a total of 133k 
readout channels. The two outermost Silicon Strip Detector (SSD) layers consist of double-sided silicon 
micro-strip sensors with 95 $\mu$m pitch, comprising a total of 2.6 million readout channels. 
The design spatial resolutions of the ITS sub-detectors ($\sigma_{r \phi} \times \sigma_{z}$) are: 
12$\times$100 $\mu$m$^{2}$ for SPD, $35 \times 25$ $\mu$m$^{2}$ for SDD, and $20 \times 830$ $\mu$m$^{2}$ for SSD. The ITS  has been aligned using reconstructed tracks from cosmic rays and from proton-proton collisions~\cite{its:alig}.

The VZERO detector consists of two forward scintillator hodoscopes. Each detector is segmented into 32 scintillator counters 
which are arranged in four rings around the beam pipe. They are located at distances $z = 3.3$ m and $z = -0.9$ m from the nominal 
interaction point and cover the pseudorapidity ranges: $2.8 < \eta < 5.1$ and $-3.7 < \eta < -1.7$, respectively. The beam-related background was rejected at offline level using the VZERO time and by cutting on the correlation between the number of clusters and track segments in the SPD.

The minimum bias (MB) trigger used in this analysis required a hit in one of the VZERO counters or in the SPD detector. In addition, a coincidence was 
required between the signals from two beam pickup counters, one on each side of the interaction region, 
indicating the presence of passing bunches~\cite{aliceNch1}.

\section{Data analysis}
MB events at $\sqrt{s}=$ 0.9 and 7 TeV (recorded in 2010) and at $\sqrt{s}=2.76$ 
TeV (recorded in 2011) have been analyzed using about $40$ million 
events, each at 7 and 2.76 TeV, and 3.6 million at 0.9 TeV. Since no energy dependence is found for the event shape observable, we present mostly results for 0.9 and 7 TeV.

The position of the interaction vertex is reconstructed by correlating hits in the two silicon-pixel layers. 
The vertex resolution depends on the track multiplicity, and is typically $0.1-0.3$ mm in the longitudinal 
($z$) and $0.2-0.5$ mm in the transverse direction. The event is accepted if its longitudinal vertex position ($z_{v}$) 
satisfies $|z_{v} - z_{0}| < 10$ cm, where $z_{0}$ is the nominal position. 

To ensure a good resolution on the transverse sphericity, only events with more than two primary tracks in $|\eta| \leq 0.8$ 
and $p_{\rm T} \geq 0.5$ GeV/$c$ are selected. The cuts on $\eta$ and $p_{\rm T}$ ensure high 
charged particle track reconstruction efficiency for primary tracks \cite{alice30}. 
These cuts reduce the available statistics to about $9.1$, $4.2$ and  $0.42$ million of MB events for the 7 TeV, 
2.76 TeV and 0.9 TeV data,  respectively. 

At 7 TeV collision energy, the fractions of non-diffrac\-tive events after the cuts are $99.5$\% and 
$93.6$\% according to PYTHIA6 version 6.421~\cite{pythia} (tune PERUGIA-0~\cite{perugia}) and PHOJET version 1.12~\cite{phojet}, 
respectively. 
In the case of single-diffractive events the fractions are $0.3$\% and $4.8$\%, while the 
double-diffractive events represent $0.2$\% and $1.6$\% of the sample as predicted by PYTHIA6 
and PHOJET, respectively.

\subsection{Track selection}
Charged particle tracks are selected in the pseudorapidity range $|\eta|\leq0.8$. 
In this range, tracks in the TPC can be reconstructed with  minimal efficiency 
losses due to detector boundaries. Additional quality requirements are applied to ensure high tracking 
resolution and low contamination from secondary and fake tracks~\cite{alice30}. 
A track is accepted if it has at least 70 space points in the TPC, and the $\chi^{2}$ per space point used for the 
momentum fit is less than 4. Tracks are rejected as not associated to the primary vertex if their distance of closest 
approach to the reconstructed event vertex in the plane perpendicular to the beam axis, $d_{0}$, exceeds
$0.245 +\frac{0.294}{p_{\rm T}^{0.9}} $  ($p_{\rm T}$ in GeV/$c$, $d_{0}$ in cm). 
This cut is tuned to select primary charged particles with high efficiency and to minimize the contributions 
from weak decays, conversions and secondary hadronic interactions in the detector material. 

\subsection{Selection of soft and hard events}
\begin{figure}[t]
\begin{center}
\includegraphics[width=0.38\textwidth]{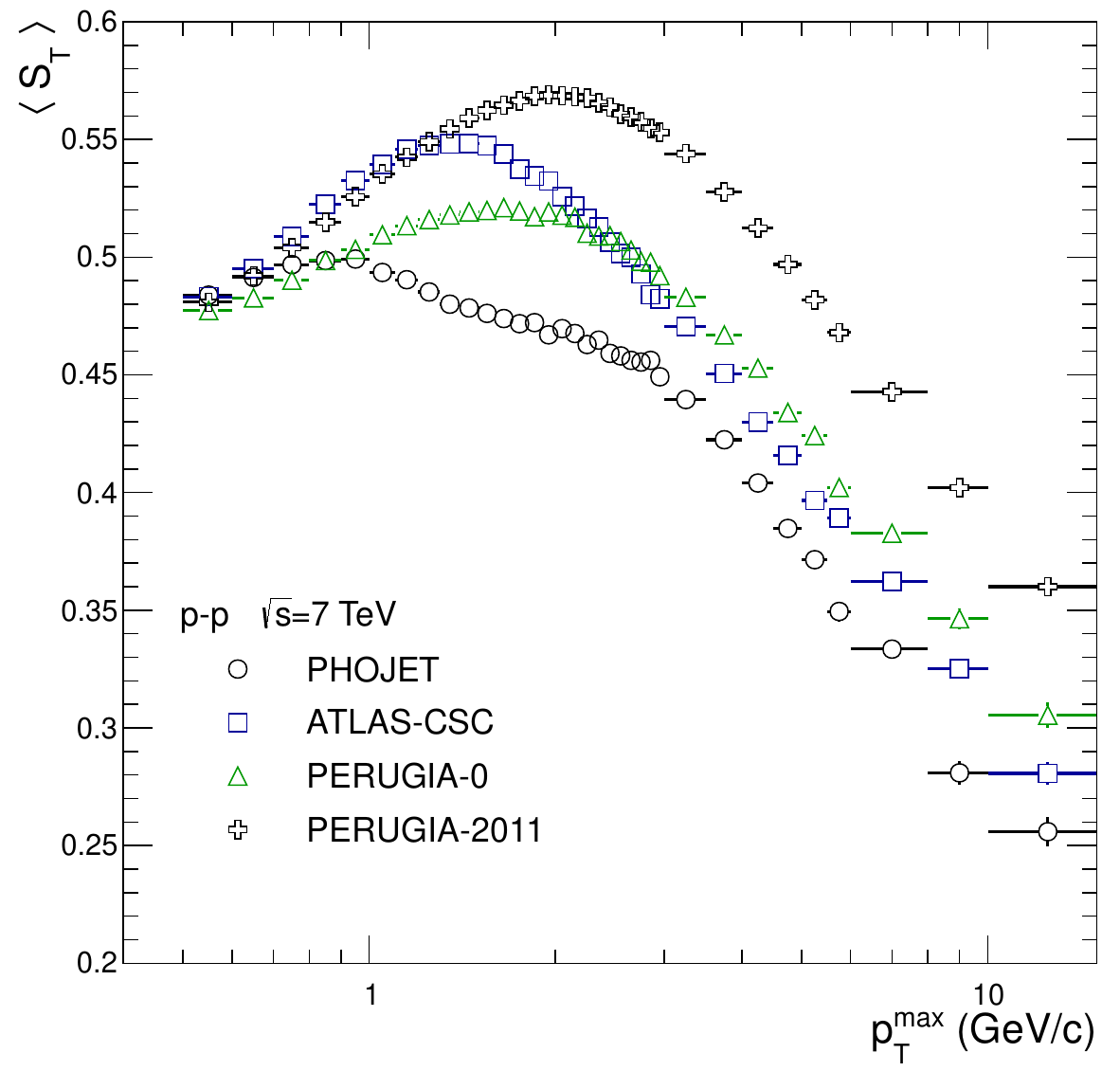}
\caption{Mean transverse sphericity versus $p_{\rm T}^{\rm max}$ for MC simulations 
at $\sqrt{s}=7$ TeV. Results are shown for PHOJET and PYTHIA6 (tunes ATLAS-CSC, PERUGIA-0 and PERUGIA-2011) simulations. The events are required to have more than 2 primary charged particles in $|\eta|\leq0.8$ 
and transverse momentum above $0.5$ GeV/$c$.}
\label{fig:1}
\end{center}
\end{figure}

The analysis is presented for two categories of events defined by the maximum charged-particle transverse momentum 
for $|\eta| \leq 0.8$ in each event. 
This method is often used in an attempt to characterize events by separating the 
different modes of production. It aims to divide the sample into two event classes: 
a) events dominantly without any hard scattering (``soft'' events) and 
b) events dominantly with at least one hard scattering (``hard'' events). Figure~\ref{fig:1} shows the mean transverse sphericity versus maximum $p_{\rm T}$ of the event obtained from  minimum bias simulations at $\sqrt{s}=7$ TeV using the particle and event cuts described previously. 
Note that PYTHIA6 simulations (tunes: ATLAS-CSC~\cite{atlascsc}, PERUGIA-0 and PERU\-GIA-2011~\cite{pe2011:a}) 
exhibit a maximum around $1.5-2.0$ GeV/$c$, while PHOJET shows an intermediate 
transition slope in $p_{\rm T}^{\rm max}=1-3$ GeV/$c$.
This observation motivated the choice of the following separation cut:  ``soft'' events are defined as events that do not have a track above 2 GeV/$c$, while ``hard'' events 
are all others. The aggregate of both classes is called ``all''. 
The selection of 2 GeV/$c$ has been motivated in the past as an accepted limit between soft and hard 
processes~\cite{hijing:1}. 
For parton-parton interactions the differential cross section is divergent for $p_{\rm T} \rightarrow 0$, 
so that a lower cut-off is generally introduced in order to regularize the 
divergence. For example in PYTHIA6, the default cut-off is $2$ GeV/$c$ for $2\rightarrow 2$ processes.

\begin{table}
\begin{center}
\begin{tabular}{l|c|c|c}
\hline
\hline
& {\bf 0.9 TeV}   & {\bf 2.76 TeV}  & {\bf 7 TeV}  \\ 
\hline
{\bf ALICE (data)} &  5.7 & 3.54 & 2.36 \\
{\bf PHOJET} &  8.53 & 4.34 & 2.52\\
{\bf ATLAS-CSC}  & 10.95 &  5.76 & 3.41\\
{\bf PERUGIA-0} &  5.6 &  3.26 & 2.06\\
{\bf PERUGIA-2011} &  6.78 & 3.64 & 2.29\\
{\bf PYTHIA8} &  7.28 & 3.92 & 2.37\\
\hline
\hline
\end{tabular}
\end{center}
\caption{Ratio of the number of ``soft'' to ``hard'' events for data and MC generators according to the 
event selection criteria defined in the text. Corrections for trigger and vertexing inefficiency have been 
applied resulting in $<2\%$ systematic uncertainty.}
\label{tab:1}
\end{table}

Table~\ref{tab:1} shows the ratio of ``soft'' to ``hard'' events for ALICE data and the generators: PHOJET, PYTHIA6 (tunes ATLAS-CSC, PERUGIA-0 and PERUGIA-2011) and PYTHIA8  version 8.145~\cite{pythia8}. It illustrates the difficulties to reproduce the evolution of simple observables with collision energy.

\begin{figure}[t]
\begin{center}
\includegraphics[width=0.4\textwidth]{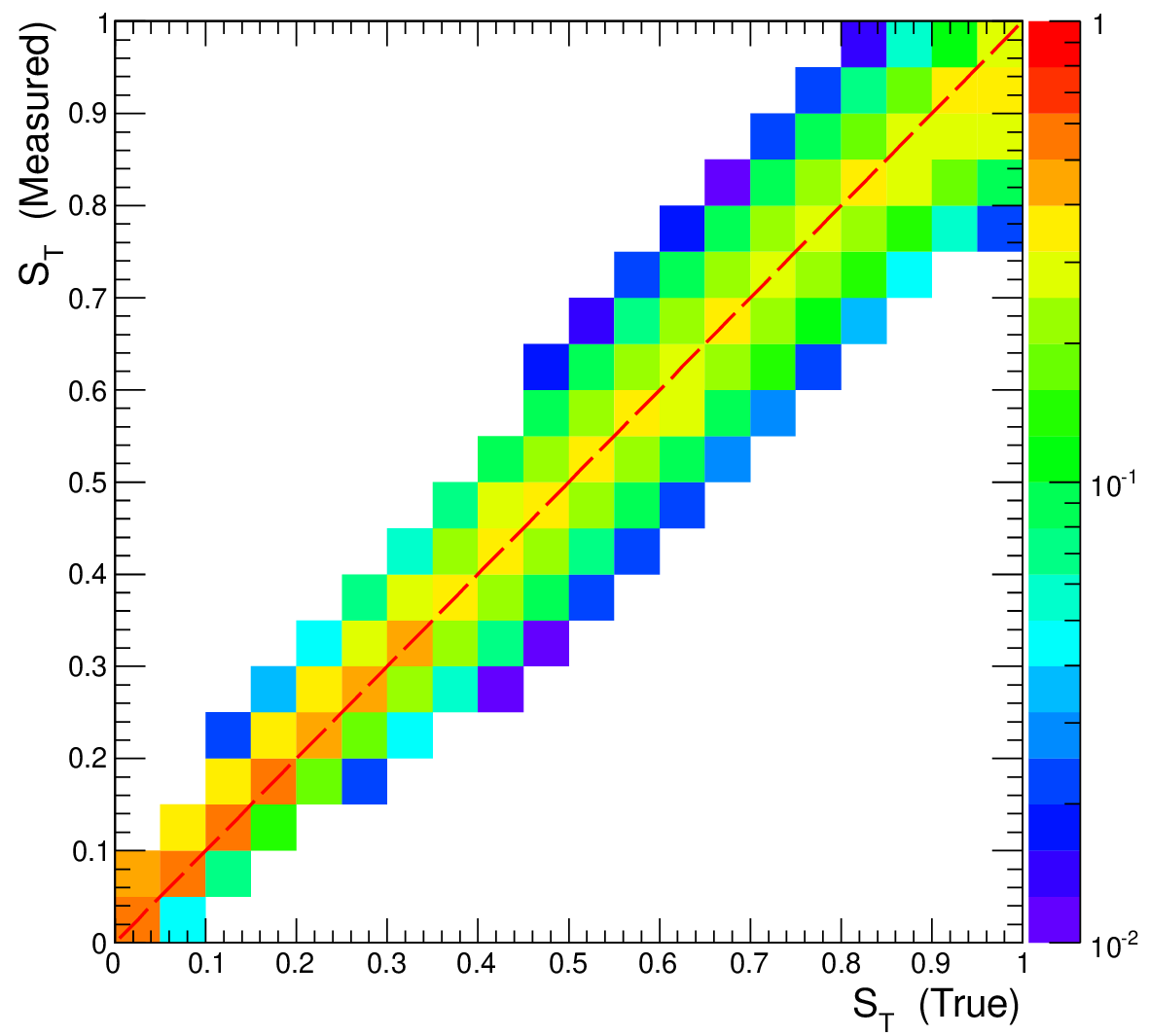}
\caption{Example of the sphericity response matrix for a measured multiplicity of 25 charged particles at mid-rapidity. The events are generated using the PYTHIA6 tune ATLAS-CSC (pp collisions at 
$\sqrt{s}=7$ TeV)  and then transported through the detector. 
Particles and tracks with $|\eta|\leq0.8$ and $p_{\rm T}\geq0.5$ GeV/$c$ are used.} 
\label{fig:2}
\end{center}
\end{figure}

\begin{figure}
\begin{center}
\includegraphics[width=0.4\textwidth]{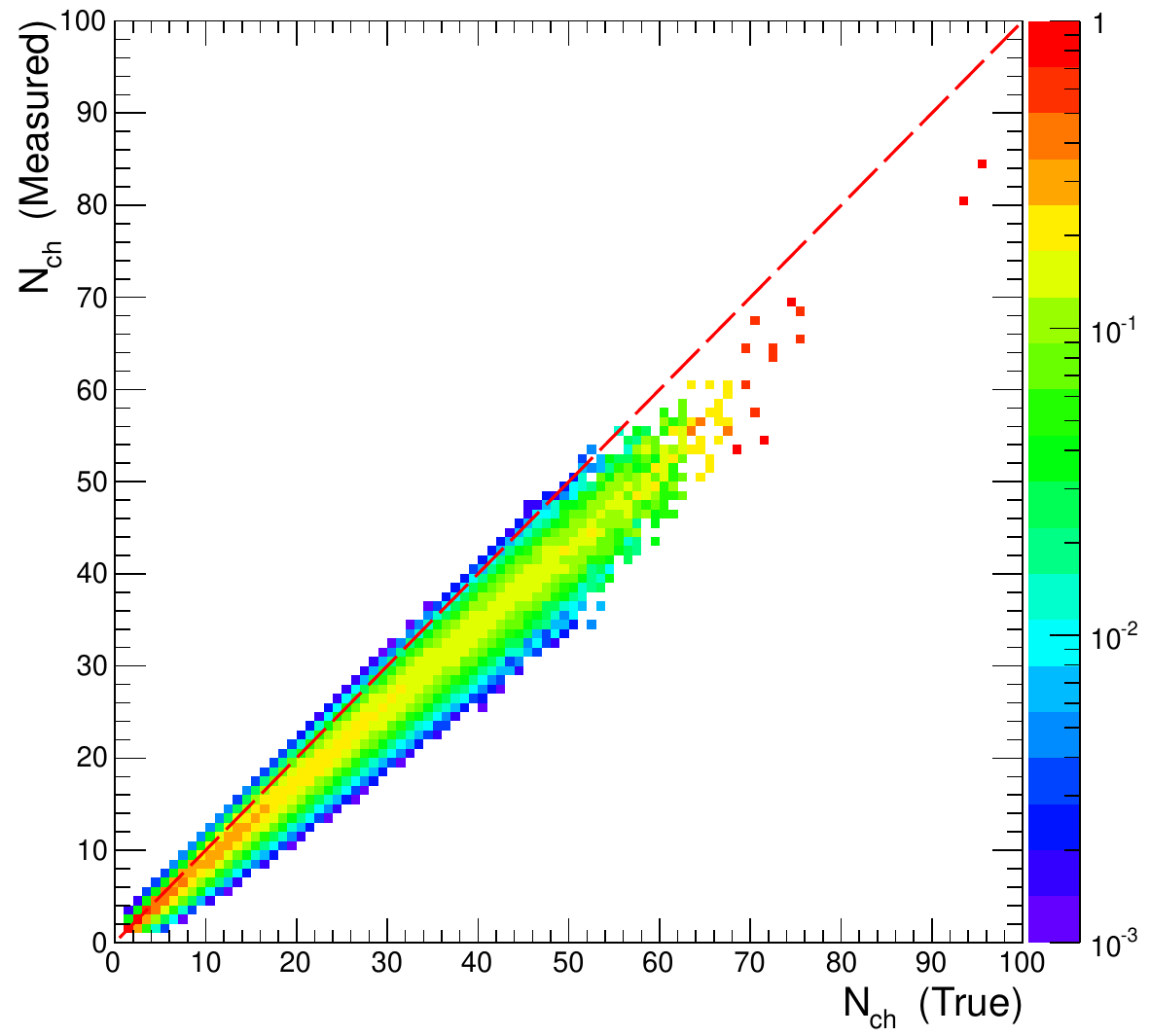}
\caption{Example of the multiplicity response matrix. The events are generated using PYTHIA6 tune 
ATLAS-CSC (pp collisions at $\sqrt{s}=7$ TeV) and then transported through 
the detector. Particles and tracks with $|\eta|\leq0.8$ and $p_{\rm T}\geq0.5$ GeV/$c$ are used. }
\label{fig:3}
\end{center}
\end{figure}

\begin{figure}[t]
\begin{center}
\includegraphics[width=0.4\textwidth]{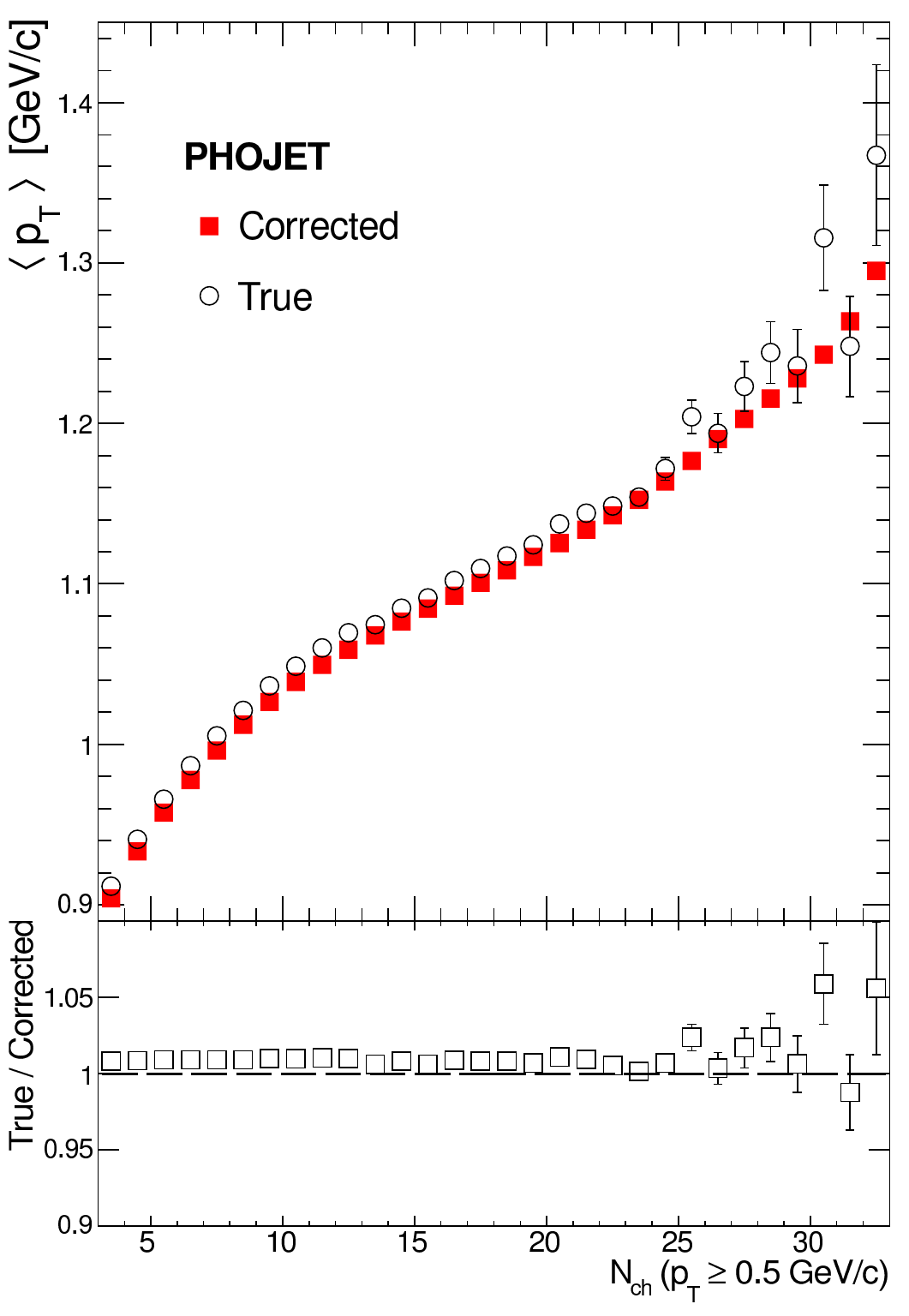}
\caption{Performance of the procedure to correct the reconstructed mean $p_{\rm T}$ as a function of multiplicity 
for ``all'' events. The method is tested using PHOJET as input and applying corrections derived from PYTHIA. 
The MC true (PHOJET result at generation level) is compared with the corrected result after simulation and 
reconstruction.} 
\label{fig:4}  
\end{center}
\end{figure}

\subsection{Corrections}
The MC simulations used to compute the correction include transport through the detector and full reconstruction 
with the same algorithms as the data. 

To correct the measured mean sphericity for efficiency, acceptance, and other detector effects, and to obtain it as the number of charged particles ($N_{\rm ch}$) in $|\eta| \leq 0.8$ two steps were followed. First, the measured sphericity distributions in bins of measured mid-rapidity charged particle multiplicity ($N_{\rm m}$) are unfolded using the detector sphericity response matrices. The unfolding implements a $\chi^{2}$ minimization with regularization~\cite{blobel:1}. Second, to account for the experimental resolution of the measured multiplicities, the mean values of the unfolded distributions  ($\left\langle S_{\rm T} \right\rangle^{\rm unf}$) are weighted by the detector multiplicity response, \\ ${R}(N_{\rm ch},N_{\rm m})$. This procedure can be seen as

\begin{equation}
\left\langle S_{\rm T} \right\rangle (N_{\rm ch}) = 
\sum_{m} \left\langle S_{\rm T} \right\rangle^{\mathtt{\rm unf}} (N_{\rm m}) R(N_{\rm ch},N_{\rm m})\,.
\end{equation}

Figures~\ref{fig:2} and~\ref{fig:3} show  an example of the sphericity response matrix with a measured multiplicity of 25 charged particles at mid-rapidity  and the multiplicity response matrix, respectively. 
The MC simulations are based on the PYTHIA6 tune ATLAS-CSC. Different simulations were tested, and all produce the same results to within $1$\%.

The sphericity distributions in four bins of multiplicity: (a) $3 \leq N_{\rm ch} < 10$, (b) $10 \leq N_{\rm ch} < 20$,  (c) $20 \leq N_{\rm ch} < 30$ and (d) $N_{\rm ch} \geq 30$ are also presented. The normalized spectra give the probability of finding an event with certain sphericity at given multiplicity. The normalized spectra were corrected bin-by-bin as follows 

\begin{equation}
P(S_{\rm T})\mid_{{\rm at} \, N_{\rm ch}} = 
P(S_{\rm T}^{\rm m})\mid_{{\rm at}\,N_{\rm m}} \times C_{1} \times C_{2}\,,
\end{equation}
where $P(S_{\rm T}^{\rm m})\mid_{{\rm at} \, N_{\rm m}}$ is the measured probability of finding an event 
with sphericity $S_{\rm T}$ in a bin of measured multiplicity ($N_{\rm m}$). This probability is corrected 
by $C_{1}$ and $C_{2}$, which are computed using MC. $C_{1}$ is the correction of the spectra at the measured 
multiplicity bin
\begin{equation} 
C_{1}=\frac{P(S_{\rm T}^{\rm unf})}{P(S_{\rm T}^{\rm m})}\mid_{{\rm at} \, N_{\rm m}}\,,
\end{equation}
and $C_{2}$ corrects the probability by the migration from high to low multiplicity
\begin{equation} 
C_{2}=\frac{P(S_{\rm T}^{\rm t})\mid_{{\rm at} \, N_{\rm ch}}}{P(S_{\rm T}^{\rm t})\mid_{{\rm at} \, N_{\rm m}}}\,.
\end{equation}

In the expressions, $P(S_{\rm T}^{\rm t})$ is the probability of finding an event with true sphericity $S_{\rm T}^{\rm t}$, where ``true'' refers to the value 
obtained at generator level.  $S_{\rm T}^{\rm t}$ and $S_{\rm T}^{\rm unf}$ are the true and unfolded sphericity distributions, 
respectively. The latter are the results of the unfolding of the simulated measurements, 
{\it i.e.} PYTHIA6 (tune PERUGIA0) corrected by PHOJET and vice versa.

\begin{table}[t]
\begin{center}
\begin{tabular}{l|ccc}
\hline
\hline
{\bf Contribution}  & {\bf All}  &  {\bf Soft} & {\bf Hard}   \\ 
\hline
Track selection cuts & 0.3$\%$ & 0.3$\%$ & 0.3$\%$  \\
Event generator dependence & 0.5$\%$ & 0.5$\%$ & 2$\%$  \\
Different run conditions & 1.0$\%$ & 1.0$\%$ & 1.0$\%$  \\
Secondary track rejection & $<0.8\%$ & $<0.8\%$ & $<0.8\%$ \\
Pile-up events & 0.2$\%$ & 0.2$\%$ & 0.2$\%$ \\
Method ($N_{\rm ch}<5$) & $<\,5.0\%$ & $<\,5.0\%$ & $<\,11.0\%$ \\
Method ($N_{\rm ch}\geq5$) & $<1.5\%$ & $<1.5\%$ & $<1.5\%$ \\
Detector misalignment  &  negl. & negl. & negl. \\
ITS efficiency  &  negl. & negl. & negl. \\
TPC efficiency  &  negl. & negl. & negl. \\
Beam-gas events  &  negl. & negl. & negl. \\
\hline
{\bf Total}  ($N_{\rm ch}<5$) & $<\,6.0\%$& $<\,6.0\%$ & $<\,12.0\%$  \\
{\bf Total}  ($N_{\rm ch}\geq 5$) & $<\,2.2\%$& $<\,2.2\%$ & $<\,3.0\%$  \\
\hline
\end{tabular}
\end{center}
\caption{Contributions to the systematic uncertainties on the mean transverse sphericity $\langle S_{\rm T} \rangle$.} 
\label{tab:2}
\end{table}

\begin{table}
\begin{center}
\begin{tabular}{l|ccc}
\hline
\hline
{\bf Multiplicity range}  & {\bf 3-9}  &  {\bf 10-19} & {\bf 20-29}   \\ 
\hline
Method & $<0.1\%$ & $<2.0\%$ & $<5.0\%$  \\
Event generator dependence & $<5.0\%$ & $<1.0\%$ & $<1.0\%$  \\
Pile-up events & $<1.0\%$ & $<1.0\%$ & $<4.0\%$ \\
\hline
{\bf Total}  & $<\,5.1\%$& $<\,2.4\%$ & $<\,6.5\%$  \\ 
\hline
\end{tabular}
\end{center}
\caption{Systematic uncertainties on the sphericity distributions.} 
\label{tab:3}
\end{table}

\begin{figure*}[t]
\begin{center}

\begin{tabular}{cc}
\includegraphics[width=0.9\textwidth]{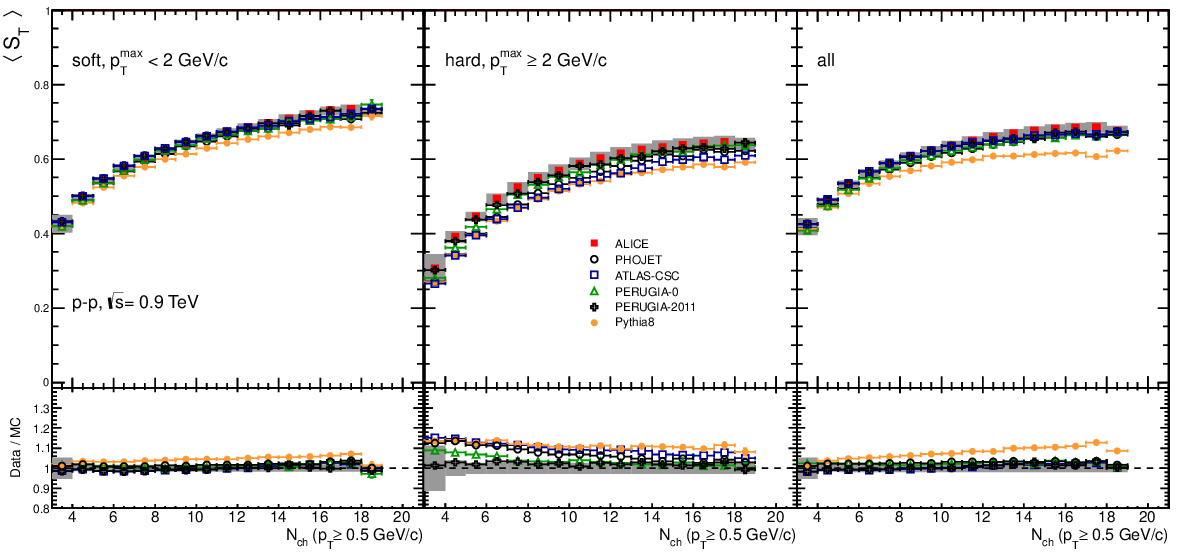} 

\\
\includegraphics[width=0.9\textwidth]{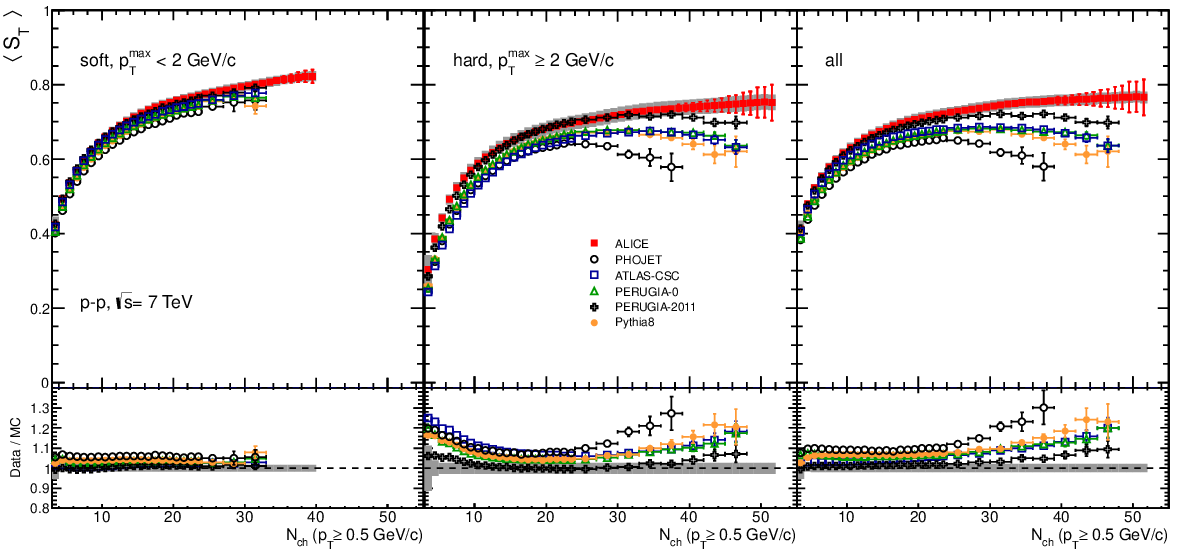} &

\\
\end{tabular}
\caption{Mean transverse sphericity as a function of charged particle multiplicity. 
The ALICE data are compared with five models: PHOJET, PYTHIA6 (tunes: ATLAS-CSC, PERUGIA-0 and PERUGIA-2011) and PYTHIA8. Results at $\sqrt{s}=0.9$ and 7 TeV are shown in the 
top and bottom rows, respectively. Different event classes are presented: (left) ``soft'', (middle) ``hard'' and (right) ``all'' 
(see text for definitions). 
The statistical errors are displayed as error bars and the systematic uncertainties 
as the shaded area. The horizontal error bars indicate the bin widths. 
Symbols for data points and model predictions are presented in the legend.}
\label{fig:5}
\end{center}
\end{figure*}

Finally, to determine  $\langle p_{\rm T} \rangle (N_{\rm ch})$, we take   the mean $p_{\rm T}$ by counting all 
tracks that pass the cuts discussed above as a function of measured multiplicity ($N_{\rm m}$). 
Once we get  $\langle p_{\rm T} \rangle^{\rm m} (N_{\rm m})$ we follow the approximation
\begin{equation}
\left\langle p_{\rm T} \right\rangle (N_{\rm ch}) = \sum_{m} \left\langle p_{\rm T}
 \right\rangle^{\rm m} (N_{\rm m}) {R}(N_{\rm ch},N_{\rm m})\,.
\end{equation}
Note that in this case an unfolding of the mean $p_{\rm T}$ is not implemented. Figure~\ref{fig:4} illustrates 
the performance of the procedure using PHOJET simulations as input. The response matrices are computed as above using the PYTHI\-A6 event generator. 
The corrected points are compared with MC at generation level. The differences, at high multiplicity, 
reach about $1.5\%$.

\begin{figure*}[t]
\begin{center}
\includegraphics[width=0.95\textwidth]{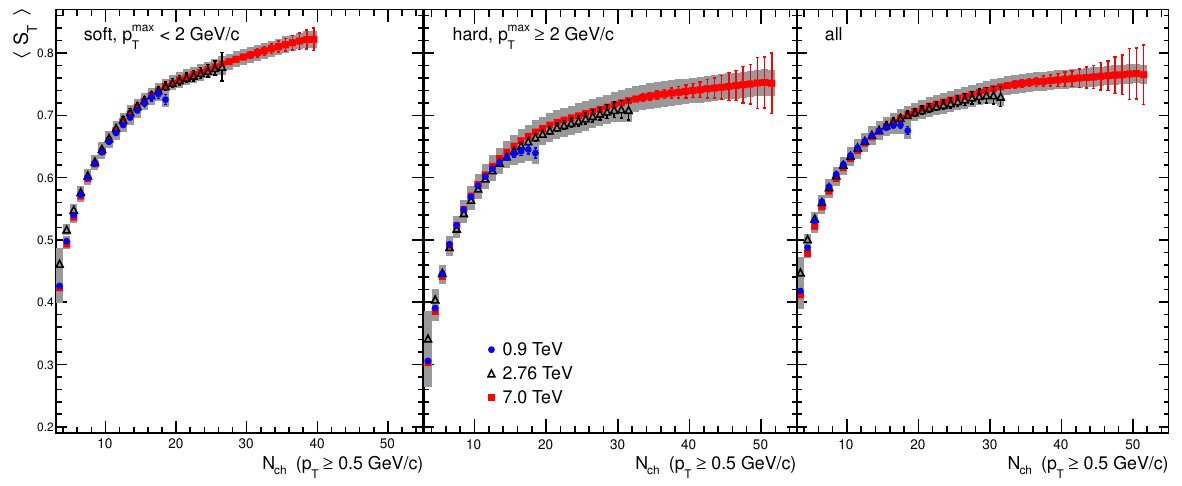}
\caption{Mean sphericity versus multiplicity for (left) ``soft'', (middle) ``hard'' and (right) ``all'' events for 
$\sqrt{s}=$ 0.9, 2.76 and 7 TeV. The statistical errors are displayed as error bars and the 
systematic uncertainties as the shaded area.}
\label{fig:6}
\end{center}
\end{figure*}

\subsection{Systematic uncertainties}
The systematic uncertainties on $\langle S_{\rm T}\rangle$ are evaluated as follows.
To minimize the adverse effects of pile-up on the multiplicity only runs with a low probability of multiple 
collisions were used. The parameter used to measure the pile-up level is the median of the Poisson 
distribution which is based on the recorded beam luminosity, it is assumed to characterize the probability  to have $n$ interactions reconstructed as a single event. 
Furthermore, a cut on the number of extra vertices reconstructed by the SPD was introduced. 
The systematic uncertainty was estimated from the differences between results using runs with the 
smallest and largest pile-up probability, and for the ``all'' sample was found to be less than 0.2\%.
The uncertainty due to the rejection of secondaries was estimated by increasing their contribution up to 
$\sim 8 \%$. This is done by varying the cut on the distance of closest approach ($d_{0}>0.0350+\frac{0.0420}{p_{\rm T}^{0.9}}$, $p_{\rm T}$ in GeV/$c$, $d_{0}$ in cm) of the
considered track to the primary vertex in the plane perpendicular to the beam. 
The event generator dependence was determined from a comparison of the results obtained when either 
PYTHIA6 or PHOJET were used to compute the correction matrices, and found to be of the order
of few \%.
The most significant contribution to the systematic uncertainties is due to the method of correction. 
It was estimated from MC by the ratio true-$S_{\rm T}$ to corrected-$S_{\rm T}$ as a function 
of multiplicity. 
For example, the largest uncertainty is at low multiplicity ($N_{\rm ch}\sim 3$) for the ``hard'' sample, 
where it reaches $\sim 11\%$.
Different sets of cuts were implemented in order to estimate the systematic uncertainty 
due to track selection.
Table~\ref{tab:2} summarizes the systematic uncertainties on $\langle S_{\rm T} \rangle$.
In addition, other checks were performed to ensure an accurate interpretation of the results. For instance, when applying the analysis to randomized events (where the track azimuthal angles are uniformly distributed between $0$ and $2\pi$), we obtain results that are about $10\%$ larger than in data. The conclusion is that measured sphericity in data is not the result of a random track combination. Also, the analysis was applied to events with sphericity axes in different regions of 
the TPC, to ensure that the results are not biased by any residual geometry effects. 

In the case of the mean transverse momentum as a function of multiplicity the systematic uncertainties 
are taken from~\cite{alice30}, the only difference being the method of correction. 
The uncertainty was estimated by applying the correction algorithm to reconstructed events generated with 
PYTHIA6, while the correction matrices were computed using events generated with PHOJET. 
The final distributions  were compared with the results at generator level. 
For the ``all'' sample the uncertainty reaches $1.5\%$, while for ``soft'' and ``hard''
it reaches $1.0\%$ and $5.1\%$, respectively.

For the case of the sphericity distributions in intervals of multiplicity, the main uncertainties are listed in 
Table~\ref{tab:3}. They were estimated following similar procedures as described above.

\begin{figure*}[t]
\begin{center}

\begin{tabular}{cc}
\includegraphics[width=0.9\textwidth]{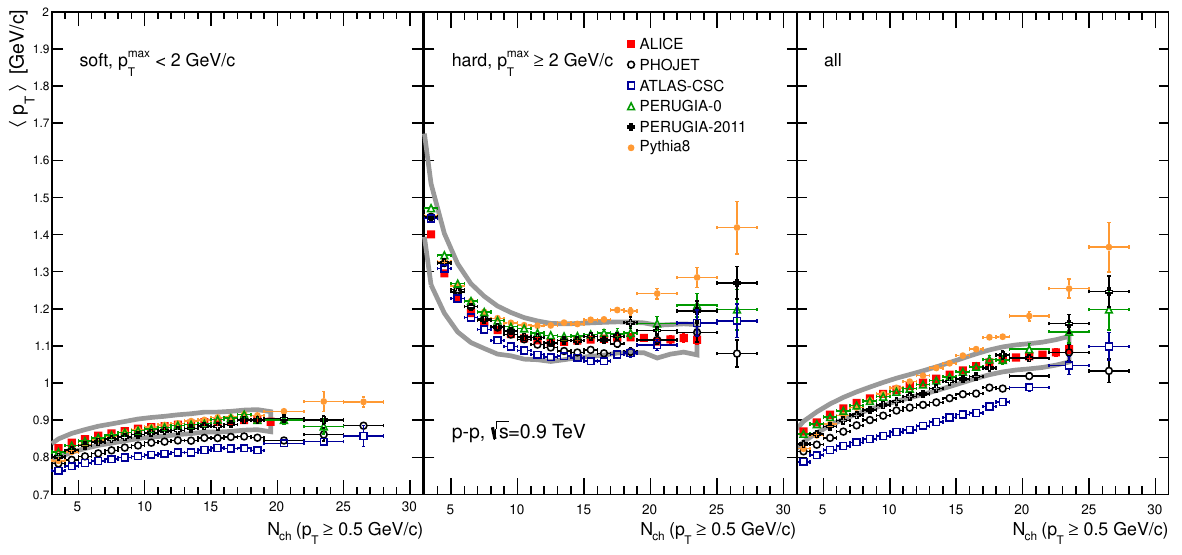} 

\\
\includegraphics[width=0.9\textwidth]{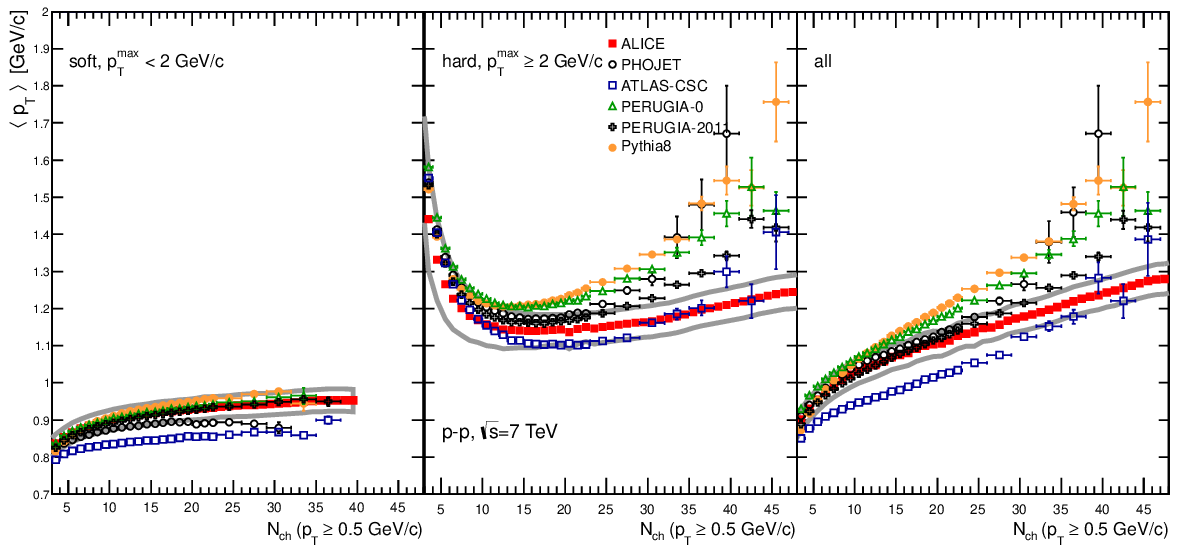} &
\\
\end{tabular}

\caption{ Mean transverse momentum versus multiplicity. The ALICE data are compared with five models: 
PHOJET, PYTHIA6 (tunes: ATLAS-CSC, PERUGIA-0 and PERUGIA-2011) and 
PYTHIA8. Results at $\sqrt{s}=0.9$ and 7 TeV are shown in the top and bottom rows, respectively. 
Different event classes are presented: (left) ``soft'', (middle) ``hard'' and (right) ``all''. The gray lines 
indicate the systematic uncertainty on data and the horizontal error bars indicate the bin widths.}
\label{fig:7}
\end{center}
\end{figure*}

\section{Results}
In this section the results of the analyses are presented along with predictions of different models: PHOJET, PY\-THIA6 version (tunes: ATLAS-CSC, PERUGIA-0 and PE\- RUGIA-2011) and PYTHIA8.

\begin{figure*}[t]
\begin{center}
\includegraphics[width=0.8\textwidth]{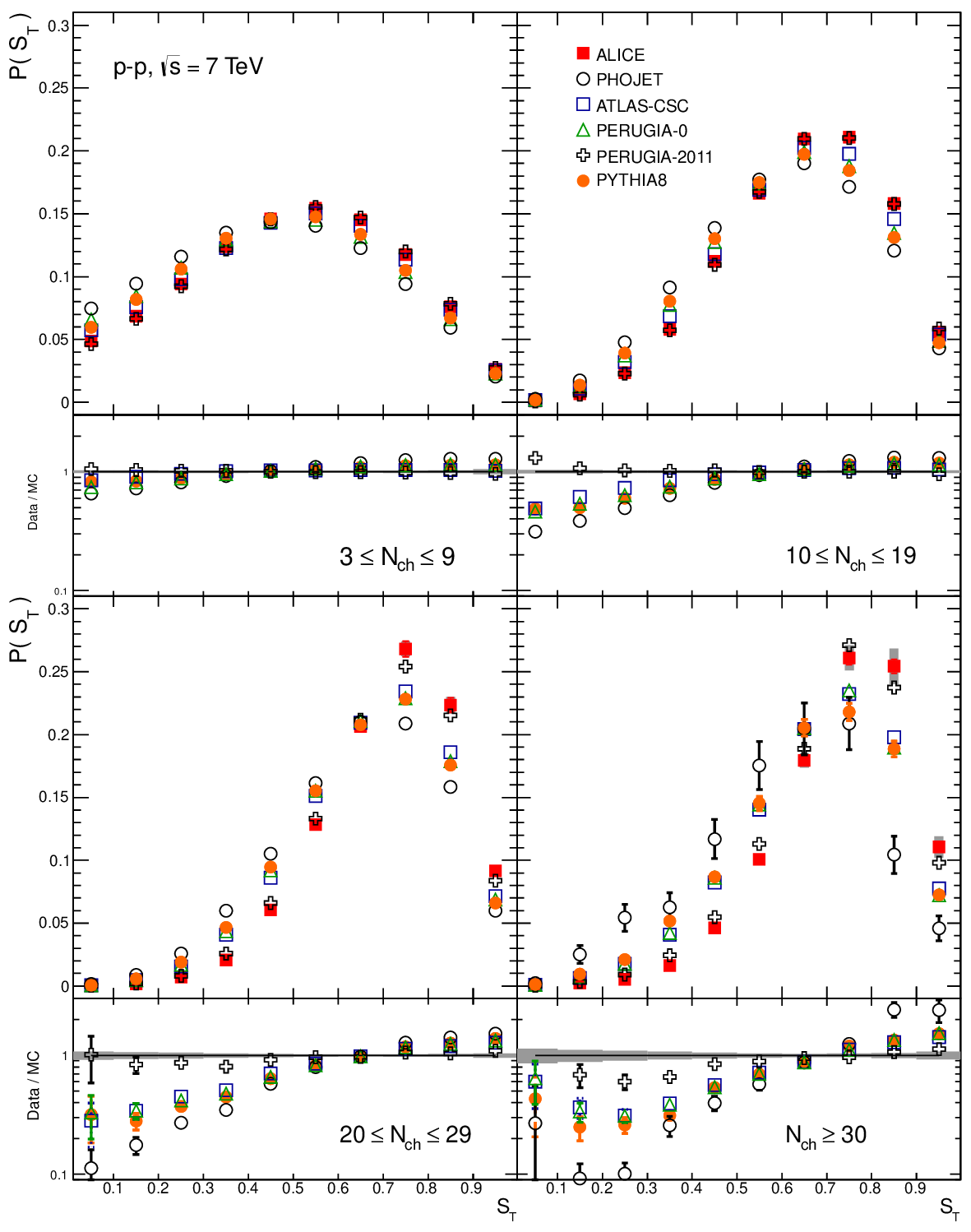}
\caption{Sphericity distributions in four bins of multiplicity: (upper-left) $3 \leq N_{\rm ch} \leq 9$, (upper-right) $10 < N_{\rm ch} \leq 19$, 
(bottom-left) $20 < N_{\rm ch} \leq 29$ and (bottom-right) $N_{\rm ch} \geq 30$ at $\sqrt{s}=7$ TeV. The statistical errors are displayed 
as error bars and the systematic uncertainties as the shaded area.}
\label{fig:8}
\end{center}
\end{figure*}

\begin{figure*}[t]
\begin{center}
\includegraphics[width=0.6\textwidth]{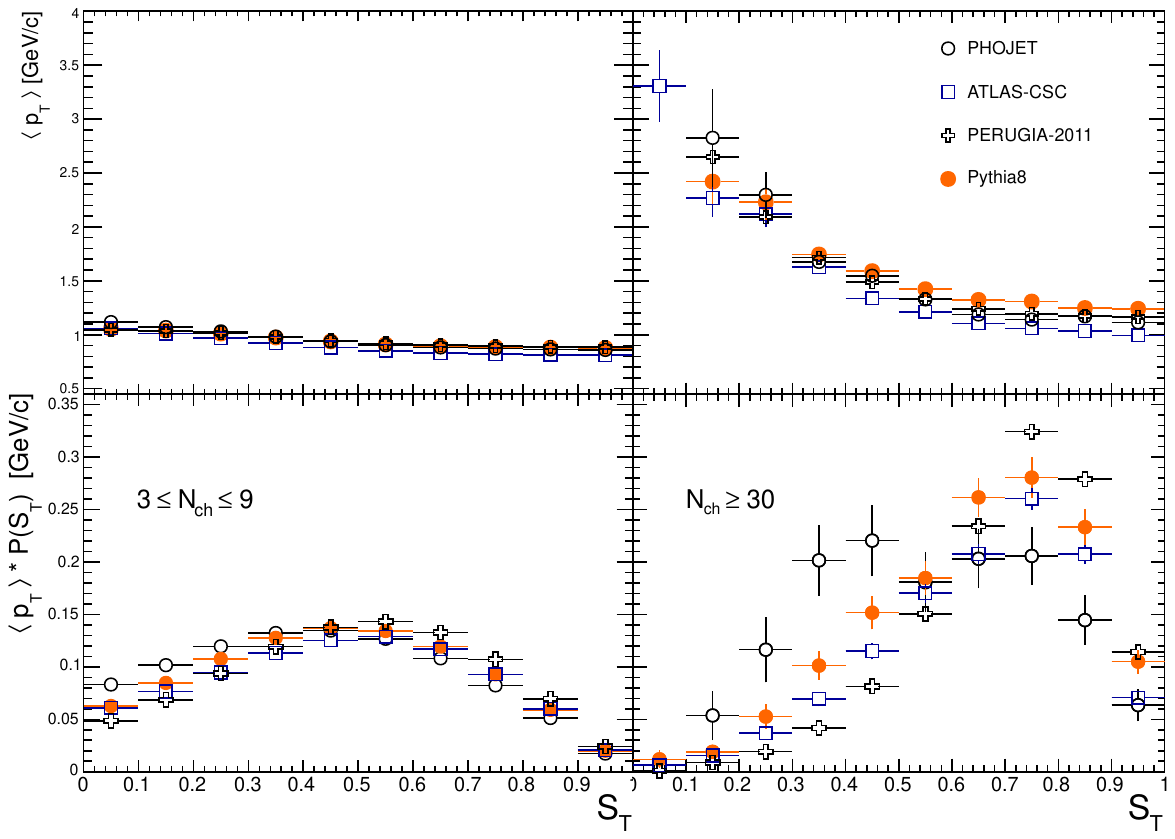}
\caption{Mean $p_{\rm T}$ (top) as a function of sphericity for two multiplicity bins 
(left) $3 \leq N_{\rm ch} \leq 9$ and (right) $N_{\rm ch} \geq30$ for minimum bias pp collisions at $\sqrt{s}=7$ TeV simulated 
with four different MC generators: PHOJET, PYTHIA6 (tune ATLAS-CSC and PERUGIA-2011) and PYTHIA8. Also the contributions of 
the different event topologies to the averaged mean $p_{\rm T}$ are presented (bottom).}
\label{fig:9}
\end{center}
\end{figure*}

\subsection{Mean sphericity}
The mean transverse sphericity as a function of $N_{\rm ch}$ at $\sqrt{s}=0.9$ and 7 TeV is shown in 
Fig.~\ref{fig:5} for the different event classes. 
The mean sphericity (right panel) increases up to around $15$ primary charged particles, however, for larger multiplicities 
the ALICE data exhibit an almost constant or slightly rising behavior.
For ``soft'' events and $\sqrt{s}=0.9$ TeV, the models are in agreement with the ALICE 
measurements over the full range of multiplicity, except for PYTHIA8 prediction,
which is $5-10$\% lower. There is insufficient statistics to perform the unfolding for $N_{ch}>18$.
At 7 TeV, the differences between models and data are below $10\%$ for ``soft'' events.
For  the ``hard'' events, PHOJET, ATLAS-CSC, PERUGIA-0 and PYTHIA8 predict a lower $\langle S_{\rm T} \rangle$ than observed in data, actually the differences between models and data are larger than $10\%$ for multiplicities below 10 and larger than 40, that is true at 0.9 and 7 TeV. The differences observed are larger than the systematic and statistical uncertainties. It is interesting to note that PERUGIA-2011 describes
the data quite well.  The fraction of ``soft'' and ``hard'' events in data and MC simulations
as a function of $N_{ch}$ (integral values given in Table~\ref{tab:1}) is found
to be different between data and the event generators.
At large $N_{ch}$, the event generators generally produce more ``hard''
events than observed in data. This difference is reflected in the
``all'' event class, since more ``hard'' events contribute in the
case of the generators, while more ``soft'' events in the case of
data. The largest isotropy in the azimuth is found at high multiplicity, $N_{\rm ch}>40$ ($|\eta| \leq 0.8$, $p_{\rm T}\geq 0.5$ GeV/$c$), in a similar multiplicity region where the CMS collaboration discovered the long-range near-side angular correlations~\cite{cms:ridge}.
Comparing the results at 0.9 and 7 TeV, it is seen that except for Pythia8 the predictions of models describe better the 0.9 TeV data than the 7 TeV ones. Lastly, the mean sphericity evolution with multiplicity at the three measured energies are shown in Fig.~\ref{fig:6} for ``soft'', ``hard'' and ``all'' events at $\sqrt{s}=$ 0.9, 2.76 and 7 TeV. The functional form of the mean sphericity as a function of $N_{\rm ch}$ is the same at all three energies in the overlapping multiplicity region.

\subsection{Mean transverse momentum}
The mean transverse momentum as a function of $N_{\rm ch}$ at $\sqrt{s}=0.9$ and 7 TeV is shown in 
Fig.~\ref{fig:7}. As seen in left panel, PERUGIA-0, PERUGIA-2011 and PYTHIA8 are within the
systematic uncertainty bands of the data for soft events, though PYTHIA8 has a different
functional form than the data. For the ``hard'' events there is a significant difference between the data and the generators 
above a multiplicity of about 20, in particular for the 7 TeV data. 
For lower multiplicities, ATLAS-CSC has an overall different shape than the other generators.
For ``all'' events, at 0.9 TeV PERUGIA-0 and PERUGIA-2011 best reproduces the data, while the rest of the models do not give 
a good description. At 7 TeV, the calculations exhibit a change in the slope around $N_{\rm ch}=30$, which is not 
observed in the data. At similar multiplicities, the MC mean sphericity reaches a maximum before it decreases with increasing multiplicity (Fig.~\ref{fig:5}). The similarity in the multiplicity dependence between 
$\langle S_{\rm T}\rangle$ and $\langle p_{\rm T}\rangle$ suggests that the models may generate more back-to-back 
correlated high $p_{T}$ particles (jets) than present in the data.

\subsection{$S_{\rm T}$ spectra in multiplicity intervals}
To disentangle the ambiguities between $p_{\rm T}$, $S_{\rm T}$ and multiplicity, 
the normalized transverse sphericity  spectra (the probability of having events of different 
transverse sphericity in a given multiplicity interval) are computed at 7 TeV for four different intervals of 
multiplicity: $N_{\rm ch}=3$--$9$, $10$--$19$, $20$--$29$ and above $30$. 
These are shown in Fig.~\ref{fig:8} along with their ratios to each MC calculation. 
In the first multiplicity bin ($N_{\rm ch}=3$--$9$), the agreement between data and MC is generally good, but in the 
second bin ($N_{\rm ch}=10$--$19$) the ratio data to MC is systematically lower for $S_{\rm T} \leq 0.4$ except for PERUGIA-2011.
In the last bin of multiplicity the overproduction of back-to-back jets (in the azimuth) reaches a factor of 3, 
and there is an underestimation of isotropic events by a factor 2. As in previous cases, the best description is done by PERUGIA-2011.

To obtain information about the interplay between multiplicity and $\langle p_{\rm T} \rangle$ through the event 
shapes, we also investigated the $\langle p_{\rm T} \rangle$ as a function of 
$\langle S_{\rm T} \rangle$ in intervals of multiplicity. 
The study is presented using MC generators at $\sqrt{s}=7$ TeV, but the conclusion also holds at the other 
two energies. Figure~\ref{fig:9} shows $\langle p_{\rm T} \rangle$ as a function of $S_{\rm T}$ for two 
multiplicity bins (top panels) along with the contribution of each sphericity bin (bottom panels) to the final 
$\langle p_{\rm T} \rangle$, {\it i.~e.} the $\langle p_{\rm T}\rangle$ weighted by the value $P(S_{\rm T})$.
There are two points to emphasize. First, a large dependence of $\langle p_{\rm T}\rangle$ on sphericity is 
observed for high multiplicities while at low ones the dependence is weaker. Second, the sphericity distribution determines the mean $p_{\rm T}$ in a specific bin of multiplicity. For instance, for $S_{\rm T} =0.3$--$0.4$
PHOJET and ATLAS-CSC have nearly the same value of $\langle p_{\rm T}\rangle$, while the contribution to 
$\langle p_{\rm T}\rangle$ in the multiplicity bin is twice larger for PHOJET compared to ATLAS-CSC.
Hence, the reproduction of the sphericity should be taking into account in the tuning of the MC generators.

\section{Conclusion}
A systematic characterization of the event shape in minimum bias proton--proton collisions at 
$\sqrt{s}=0.9$, $2.76$ and $7$ TeV is presented. Confronted with the persistent difficulties of event 
generators to reproduce simultaneously the charged particle transverse momentum and multiplicity, 
the transverse sphericity is used to provide insight into the particle production mechanisms.
The observables are measured using primary charged tracks with $p_{\rm T}\geq0.5$ GeV/$c$ in $|\eta|\leq0.8$ 
and reported as a function of the charged particle multiplicity at mid-rapidity ($N_{\rm ch}$) for events 
with different scales (``soft'' and ``hard'') defined by the transverse momentum of the leading particle. 
The data are compared with calculations of standard Monte Carlo event generators: PHOJET, PYTHIA6 (tunes: ATLAS-CSC, PERUGIA-0 and PERUGIA-2011) and PYTHIA8 (default MB parameters).

The MC generators exhibit a decrease of $\langle S_{\rm T} \rangle$ at high multiplicity with a simultaneous 
steep rise of $\langle p_{\rm T} \rangle$. On the contrary, in ALICE data $\langle S_{\rm T} \rangle$ stays 
approximately constant or slightly rising (Fig.~\ref{fig:5}) accompanied with a mild increase in 
$\langle p_{\rm T} \rangle$ (Fig.~\ref{fig:7}). 
The mean sphericity seems to primarily depend on the multiplicity and not on $\sqrt{s}$ (Fig.~\ref{fig:6}). 
At high multiplicity ($N_{\rm ch} \geq 30$) the generators underestimate the production of isotropic events and 
overestimate the production of pencil-like events (Fig.~\ref{fig:8}). It seems that the generators tend to produce large multiplicity events by favoring the production 
of back-to-back high-$p_{\rm T}$ jets (low $S_{\rm T}$) more so than in nature. 
The level of disagreement between data and generators is markedly different for ``soft'' and ``hard'' events,
being much larger for the latter~(Figs.~\ref{fig:5}-\ref{fig:7}). It is worthwhile to point out that
PERUGIA-2011 describes the various aspects of the data generally quite well, except for the mean $p_{T}$, which it overestimates at high multiplicities. Our studies suggest that the tuning of generators should include the sphericity as an additional 
reference.

%
\newenvironment{acknowledgement}{\relax}{\relax}
\begin{acknowledgement}
\section{Acknowledgements}
The ALICE collaboration would like to thank all its engineers and technicians for their invaluable contributions to the construction of the experiment and the CERN accelerator teams for the outstanding performance of the LHC complex.
\\
The ALICE collaboration acknowledges the following funding agencies for their support in building and
running the ALICE detector:
 \\
Calouste Gulbenkian Foundation from Lisbon and Swiss Fonds Kidagan, Armenia;
 \\
Conselho Nacional de Desenvolvimento Cient\'{\i}fico e Tecnol\'{o}gico (CNPq), 
\\
Financiadora de Estudos e Projetos (FINEP),
Funda\c{c}\~{a}o de Amparo \`{a} Pesquisa do Estado de S\~{a}o Paulo (FAPESP);
 \\
National Natural Science Foundation of China (NSFC), the Chinese Ministry of Education (CMOE)
and the Ministry of Science and Technology of China (MSTC);
 \\
Ministry of Education and Youth of the Czech Republic;
 \\
Danish Natural Science Research Council, the Carlsberg Foundation and the Danish National Research Foundation;
 \\
The European Research Council under the European Community's Seventh Framework Programme;
 \\
Helsinki Institute of Physics and the Academy of Finland;
 \\
French CNRS-IN2P3, the `Region Pays de Loire', `Region Alsace', `Region Auvergne' and CEA, France;
 \\
German BMBF and the Helmholtz Association;
\\
General Secretariat for Research and Technology, Ministry of
Development, Greece;
\\
Hungarian OTKA and National Office for Research and Technology (NKTH);
 \\
Department of Atomic Energy and Department of Science and Technology of the Government of India;
 \\
Istituto Nazionale di Fisica Nucleare (INFN) of Italy;
 \\
MEXT Grant-in-Aid for Specially Promoted Research, Ja\-pan;
 \\
Joint Institute for Nuclear Research, Dubna;
 \\
National Research Foundation of Korea (NRF);
 \\
CONACYT, DGAPA, M\'{e}xico, ALFA-EC and the HELEN Program (High-Energy physics Latin-American--European Network);
 \\
Stichting voor Fundamenteel Onderzoek der Materie (FOM) and the Nederlandse Organisatie voor Wetenschappelijk Onderzoek (NWO), Netherlands;
 \\
Research Council of Norway (NFR);
 \\
Polish Ministry of Science and Higher Education;
 \\
National Authority for Scientific Research - NASR (Autoritatea Na\c{t}ional\u{a} pentru Cercetare \c{S}tiin\c{t}ific\u{a} - ANCS);
 \\
Federal Agency of Science of the Ministry of Education and Science of Russian Federation, International Science and
Technology Center, Russian Academy of Sciences, Russian Federal Agency of Atomic Energy, Russian Federal Agency for Science and Innovations and CERN-INTAS;
 \\
Ministry of Education of Slovakia;
 \\
Department of Science and Technology, South Africa;
 \\
CIEMAT, EELA, Ministerio de Educaci\'{o}n y Ciencia of Spain, Xunta de Galicia (Conseller\'{\i}a de Educaci\'{o}n),
CEA\-DEN, Cubaenerg\'{\i}a, Cuba, and IAEA (International Atomic Energy Agency);
 \\
Swedish Research Council (VR) and Knut $\&$ Alice Wallenberg
Foundation (KAW);
 \\
Ukraine Ministry of Education and Science;
 \\
United Kingdom Science and Technology Facilities Council (STFC);
 \\
The United States Department of Energy, the United States National
Science Foundation, the State of Texas, and the State of Ohio.

\end{acknowledgement}
\newpage
%
%
\appendix
\section{The ALICE Collaboration}
\label{app:collab}

\begingroup
\small
\begin{flushleft}
B.~Abelev\Irefn{org1234}\And
J.~Adam\Irefn{org1274}\And
D.~Adamov\'{a}\Irefn{org1283}\And
A.M.~Adare\Irefn{org1260}\And
M.M.~Aggarwal\Irefn{org1157}\And
G.~Aglieri~Rinella\Irefn{org1192}\And
A.G.~Agocs\Irefn{org1143}\And
A.~Agostinelli\Irefn{org1132}\And
S.~Aguilar~Salazar\Irefn{org1247}\And
Z.~Ahammed\Irefn{org1225}
% ...
\renewcommand\labelenumi{\textsuperscript{\theenumi}~}
\section*{Affiliation notes}
\renewcommand\theenumi{\roman{enumi}}
\begin{Authlist}
\item \Adef{M.V.Lomonosov Moscow State University, D.V.Skobeltsyn Institute of Nuclear Physics, Moscow, Russia}Also at: M.V.Lomonosov Moscow State University, D.V.Skobeltsyn Institute of Nuclear Physics, Moscow, Russia
\item \Adef{Institute of Nuclear Sciences, Belgrade, Serbia}Also at: "Vin\v{c}a" Institute of Nuclear Sciences, Belgrade, Serbia
\end{Authlist}
\section*{Collaboration Institutes}
\renewcommand\theenumi{\arabic{enumi}~}
\begin{Authlist}
\item \Idef{org1279}Benem\'{e}rita Universidad Aut\'{o}noma de Puebla, Puebla, Mexico
\item \Idef{org1220}Bogolyubov Institute for Theoretical Physics, Kiev, Ukraine
\item \Idef{org1262}Budker Institute for Nuclear Physics, Novosibirsk, Russia
\item \Idef{org1292}California Polytechnic State University, San Luis Obispo, California, United States
\item \Idef{org14939}Centre de Calcul de l'IN2P3, Villeurbanne, France
\item \Idef{org1197}Centro de Aplicaciones Tecnol\'{o}gicas y Desarrollo Nuclear (CEADEN), Havana, Cuba
\item \Idef{org1242}Centro de Investigaciones Energ\'{e}ticas Medioambientales y Tecnol\'{o}gicas (CIEMAT), Madrid, Spain
\item \Idef{org1244}Centro de Investigaci\'{o}n y de Estudios Avanzados (CINVESTAV), Mexico City and M\'{e}rida, Mexico
\item \Idef{org1335}Centro Fermi -- Centro Studi e Ricerche e Museo Storico della Fisica ``Enrico Fermi'', Rome, Italy
\item \Idef{org17347}Chicago State University, Chicago, United States
\item \Idef{org1288}Commissariat \`{a} l'Energie Atomique, IRFU, Saclay, France
\item \Idef{org1294}Departamento de F\'{\i}sica de Part\'{\i}culas and IGFAE, Universidad de Santiago de Compostela, Santiago de Compostela, Spain
\item \Idef{org1106}Department of Physics Aligarh Muslim University, Aligarh, India
\item \Idef{org1121}Department of Physics and Technology, University of Bergen, Bergen, Norway
\item \Idef{org1162}Department of Physics, Ohio State University, Columbus, Ohio, United States
\item \Idef{org1300}Department of Physics, Sejong University, Seoul, South Korea
\item \Idef{org1268}Department of Physics, University of Oslo, Oslo, Norway
\item \Idef{org1145}Dipartimento di Fisica dell'Universit\`{a} and Sezione INFN, Cagliari, Italy
\item \Idef{org1270}Dipartimento di Fisica dell'Universit\`{a} and Sezione INFN, Padova, Italy
\item \Idef{org1315}Dipartimento di Fisica dell'Universit\`{a} and Sezione INFN, Trieste, Italy
\item \Idef{org1132}Dipartimento di Fisica dell'Universit\`{a} and Sezione INFN, Bologna, Italy
\item \Idef{org1285}Dipartimento di Fisica dell'Universit\`{a} `La Sapienza' and Sezione INFN, Rome, Italy
\item \Idef{org1154}Dipartimento di Fisica e Astronomia dell'Universit\`{a} and Sezione INFN, Catania, Italy
\item \Idef{org1290}Dipartimento di Fisica `E.R.~Caianiello' dell'Universit\`{a} and Gruppo Collegato INFN, Salerno, Italy
\item \Idef{org1312}Dipartimento di Fisica Sperimentale dell'Universit\`{a} and Sezione INFN, Turin, Italy
\item \Idef{org1103}Dipartimento di Scienze e Tecnologie Avanzate dell'Universit\`{a} del Piemonte Orientale and Gruppo Collegato INFN, Alessandria, Italy
\item \Idef{org1114}Dipartimento Interateneo di Fisica `M.~Merlin' and Sezione INFN, Bari, Italy
\item \Idef{org1237}Division of Experimental High Energy Physics, University of Lund, Lund, Sweden
\item \Idef{org1192}European Organization for Nuclear Research (CERN), Geneva, Switzerland
\item \Idef{org1227}Fachhochschule K\"{o}ln, K\"{o}ln, Germany
\item \Idef{org1122}Faculty of Engineering, Bergen University College, Bergen, Norway
\item \Idef{org1136}Faculty of Mathematics, Physics and Informatics, Comenius University, Bratislava, Slovakia
\item \Idef{org1274}Faculty of Nuclear Sciences and Physical Engineering, Czech Technical University in Prague, Prague, Czech Republic
\item \Idef{org1229}Faculty of Science, P.J.~\v{S}af\'{a}rik University, Ko\v{s}ice, Slovakia
\item \Idef{org1184}Frankfurt Institute for Advanced Studies, Johann Wolfgang Goethe-Universit\"{a}t Frankfurt, Frankfurt, Germany
\item \Idef{org1215}Gangneung-Wonju National University, Gangneung, South Korea
\item \Idef{org1212}Helsinki Institute of Physics (HIP) and University of Jyv\"{a}skyl\"{a}, Jyv\"{a}skyl\"{a}, Finland
\item \Idef{org1203}Hiroshima University, Hiroshima, Japan
\item \Idef{org1329}Hua-Zhong Normal University, Wuhan, China
\item \Idef{org1254}Indian Institute of Technology, Mumbai, India
\item \Idef{org36378}Indian Institute of Technology Indore (IIT), Indore, India
\item \Idef{org1266}Institut de Physique Nucl\'{e}aire d'Orsay (IPNO), Universit\'{e} Paris-Sud, CNRS-IN2P3, Orsay, France
\item \Idef{org1277}Institute for High Energy Physics, Protvino, Russia
\item \Idef{org1249}Institute for Nuclear Research, Academy of Sciences, Moscow, Russia
\item \Idef{org1320}Nikhef, National Institute for Subatomic Physics and Institute for Subatomic Physics of Utrecht University, Utrecht, Netherlands
\item \Idef{org1250}Institute for Theoretical and Experimental Physics, Moscow, Russia
\item \Idef{org1230}Institute of Experimental Physics, Slovak Academy of Sciences, Ko\v{s}ice, Slovakia
\item \Idef{org1127}Institute of Physics, Bhubaneswar, India
\item \Idef{org1275}Institute of Physics, Academy of Sciences of the Czech Republic, Prague, Czech Republic
\item \Idef{org1139}Institute of Space Sciences (ISS), Bucharest, Romania
\item \Idef{org27399}Institut f\"{u}r Informatik, Johann Wolfgang Goethe-Universit\"{a}t Frankfurt, Frankfurt, Germany
\item \Idef{org1185}Institut f\"{u}r Kernphysik, Johann Wolfgang Goethe-Universit\"{a}t Frankfurt, Frankfurt, Germany
\item \Idef{org1177}Institut f\"{u}r Kernphysik, Technische Universit\"{a}t Darmstadt, Darmstadt, Germany
\item \Idef{org1256}Institut f\"{u}r Kernphysik, Westf\"{a}lische Wilhelms-Universit\"{a}t M\"{u}nster, M\"{u}nster, Germany
\item \Idef{org1246}Instituto de Ciencias Nucleares, Universidad Nacional Aut\'{o}noma de M\'{e}xico, Mexico City, Mexico
\item \Idef{org1247}Instituto de F\'{\i}sica, Universidad Nacional Aut\'{o}noma de M\'{e}xico, Mexico City, Mexico
\item \Idef{org23333}Institut of Theoretical Physics, University of Wroclaw
\item \Idef{org1308}Institut Pluridisciplinaire Hubert Curien (IPHC), Universit\'{e} de Strasbourg, CNRS-IN2P3, Strasbourg, France
\item \Idef{org1182}Joint Institute for Nuclear Research (JINR), Dubna, Russia
\item \Idef{org1143}KFKI Research Institute for Particle and Nuclear Physics, Hungarian Academy of Sciences, Budapest, Hungary
\item \Idef{org1199}Kirchhoff-Institut f\"{u}r Physik, Ruprecht-Karls-Universit\"{a}t Heidelberg, Heidelberg, Germany
\item \Idef{org20954}Korea Institute of Science and Technology Information, Daejeon, South Korea
\item \Idef{org1160}Laboratoire de Physique Corpusculaire (LPC), Clermont Universit\'{e}, Universit\'{e} Blaise Pascal, CNRS--IN2P3, Clermont-Ferrand, France
\item \Idef{org1194}Laboratoire de Physique Subatomique et de Cosmologie (LPSC), Universit\'{e} Joseph Fourier, CNRS-IN2P3, Institut Polytechnique de Grenoble, Grenoble, France
\item \Idef{org1187}Laboratori Nazionali di Frascati, INFN, Frascati, Italy
\item \Idef{org1232}Laboratori Nazionali di Legnaro, INFN, Legnaro, Italy
\item \Idef{org1125}Lawrence Berkeley National Laboratory, Berkeley, California, United States
\item \Idef{org1234}Lawrence Livermore National Laboratory, Livermore, California, United States
\item \Idef{org1251}Moscow Engineering Physics Institute, Moscow, Russia
\item \Idef{org1140}National Institute for Physics and Nuclear Engineering, Bucharest, Romania
\item \Idef{org1165}Niels Bohr Institute, University of Copenhagen, Copenhagen, Denmark
\item \Idef{org1109}Nikhef, National Institute for Subatomic Physics, Amsterdam, Netherlands
\item \Idef{org1283}Nuclear Physics Institute, Academy of Sciences of the Czech Republic, \v{R}e\v{z} u Prahy, Czech Republic
\item \Idef{org1264}Oak Ridge National Laboratory, Oak Ridge, Tennessee, United States
\item \Idef{org1189}Petersburg Nuclear Physics Institute, Gatchina, Russia
\item \Idef{org1170}Physics Department, Creighton University, Omaha, Nebraska, United States
\item \Idef{org1157}Physics Department, Panjab University, Chandigarh, India
\item \Idef{org1112}Physics Department, University of Athens, Athens, Greece
\item \Idef{org1152}Physics Department, University of Cape Town, iThemba LABS, Cape Town, South Africa
\item \Idef{org1209}Physics Department, University of Jammu, Jammu, India
\item \Idef{org1207}Physics Department, University of Rajasthan, Jaipur, India
\item \Idef{org1200}Physikalisches Institut, Ruprecht-Karls-Universit\"{a}t Heidelberg, Heidelberg, Germany
\item \Idef{org1325}Purdue University, West Lafayette, Indiana, United States
\item \Idef{org1281}Pusan National University, Pusan, South Korea
\item \Idef{org1176}Research Division and ExtreMe Matter Institute EMMI, GSI Helmholtzzentrum f\"ur Schwerionenforschung, Darmstadt, Germany
\item \Idef{org1334}Rudjer Bo\v{s}kovi\'{c} Institute, Zagreb, Croatia
\item \Idef{org1298}Russian Federal Nuclear Center (VNIIEF), Sarov, Russia
\item \Idef{org1252}Russian Research Centre Kurchatov Institute, Moscow, Russia
\item \Idef{org1224}Saha Institute of Nuclear Physics, Kolkata, India
\item \Idef{org1130}School of Physics and Astronomy, University of Birmingham, Birmingham, United Kingdom
\item \Idef{org1338}Secci\'{o}n F\'{\i}sica, Departamento de Ciencias, Pontificia Universidad Cat\'{o}lica del Per\'{u}, Lima, Peru
\item \Idef{org1316}Sezione INFN, Trieste, Italy
\item \Idef{org1271}Sezione INFN, Padova, Italy
\item \Idef{org1313}Sezione INFN, Turin, Italy
\item \Idef{org1286}Sezione INFN, Rome, Italy
\item \Idef{org1146}Sezione INFN, Cagliari, Italy
\item \Idef{org1133}Sezione INFN, Bologna, Italy
\item \Idef{org1115}Sezione INFN, Bari, Italy
\item \Idef{org1155}Sezione INFN, Catania, Italy
\item \Idef{org1322}Soltan Institute for Nuclear Studies, Warsaw, Poland
\item \Idef{org36377}Nuclear Physics Group, STFC Daresbury Laboratory, Daresbury, United Kingdom
\item \Idef{org1258}SUBATECH, Ecole des Mines de Nantes, Universit\'{e} de Nantes, CNRS-IN2P3, Nantes, France
\item \Idef{org1304}Technical University of Split FESB, Split, Croatia
\item \Idef{org1168}The Henryk Niewodniczanski Institute of Nuclear Physics, Polish Academy of Sciences, Cracow, Poland
\item \Idef{org17361}The University of Texas at Austin, Physics Department, Austin, TX, United States
\item \Idef{org1173}Universidad Aut\'{o}noma de Sinaloa, Culiac\'{a}n, Mexico
\item \Idef{org1296}Universidade de S\~{a}o Paulo (USP), S\~{a}o Paulo, Brazil
\item \Idef{org1149}Universidade Estadual de Campinas (UNICAMP), Campinas, Brazil
\item \Idef{org1239}Universit\'{e} de Lyon, Universit\'{e} Lyon 1, CNRS/IN2P3, IPN-Lyon, Villeurbanne, France
\item \Idef{org1205}University of Houston, Houston, Texas, United States
\item \Idef{org20371}University of Technology and Austrian Academy of Sciences, Vienna, Austria
\item \Idef{org1222}University of Tennessee, Knoxville, Tennessee, United States
\item \Idef{org1310}University of Tokyo, Tokyo, Japan
\item \Idef{org1318}University of Tsukuba, Tsukuba, Japan
\item \Idef{org21360}Eberhard Karls Universit\"{a}t T\"{u}bingen, T\"{u}bingen, Germany
\item \Idef{org1225}Variable Energy Cyclotron Centre, Kolkata, India
\item \Idef{org1306}V.~Fock Institute for Physics, St. Petersburg State University, St. Petersburg, Russia
\item \Idef{org1323}Warsaw University of Technology, Warsaw, Poland
\item \Idef{org1179}Wayne State University, Detroit, Michigan, United States
\item \Idef{org1260}Yale University, New Haven, Connecticut, United States
\item \Idef{org1332}Yerevan Physics Institute, Yerevan, Armenia
\item \Idef{org15649}Yildiz Technical University, Istanbul, Turkey
\item \Idef{org1301}Yonsei University, Seoul, South Korea
\item \Idef{org1327}Zentrum f\"{u}r Technologietransfer und Telekommunikation (ZTT), Fachhochschule Worms, Worms, Germany
\end{Authlist}
\endgroup


\end{document}